\documentstyle[psfig,epsfig]{mn}

\newcommand{\be}{\begin{equation}}
\newcommand{\ee}{\end{equation}}
\newcommand{\ba}{\begin{eqnarray}}
\newcommand{\ea}{\end{eqnarray}}

\newcommand{\nn}{\nonumber \\}

\newcommand{\C}{\mbox{\boldmath $C$}}

\newcommand{\n}{\mbox{\boldmath $n$}}

\newcommand{\Tr}{{\rm Tr}\,}

\newcommand{\mnras}{MNRAS}

\def\gs{\mathrel{\raise1.16pt\hbox{$>$}\kern-7.0pt %         
\lower3.06pt\hbox{{$\scriptstyle \sim$}}}}         %         
\def\ls{\mathrel{\raise1.16pt\hbox{$<$}\kern-7.0pt %        
\lower3.06pt\hbox{{$\scriptstyle \sim$}}}}         %   

\title[Path-integral Evidence]{Path-integral Evidence}

\author[Kitching \& Taylor]
       {T. D. Kitching$^1$\thanks{t.kitching@ucl.ac.uk} \& A. N. Taylor$^2$\\
$^1$Mullard Space Science Laboratory, University College London, Holmbury St Mary, Dorking, Surrey RH5 6NT, UK\\
$^2$SUPA, Institute for Astronomy, University of Edinburgh, Royal Observatory, Blackford Hill, Edinburgh, EH9 3HJ, UK}

\date{}

\pagerange{\pageref{firstpage}--\pageref{lastpage}}

\pubyear{2015}

\begin{document}

\maketitle

\label{firstpage}

\begin{abstract}
Here we present a Bayesian formalism for the goodness-of-fit that is the 
evidence for a fixed functional form over the evidence for all functions 
that are a general perturbation about this form. 
This is done under the 
assumption that the statistical properties of the data 
can be modelled by a multivariate Gaussian distribution. 
We use this to show how one can optimise an
experiment to find evidence for a fixed function over perturbations
about this function. 
We apply this formalism to an illustrative problem of measuring perturbations in the dark energy
equation of state about a cosmological constant. 
\end{abstract}


\vspace{-1.7cm}
\section{Introduction}
Finding an objective Bayesian measure for a goodness of fit of a
function to some data is problematic because when one calculates a Bayes
factor (the ratio of evidences for particular models) the
assumption of at least two models must be made. To circumvent this
assumption one can ask what the evidence for a particular (singular)
model is over all possible perturbations about that model. We present
such an evidence ratio here, under some assumptions, 
which is calculated using a path-integral
methodology (an extension of the formalism described in Taylor \&
Kitching, 2010 and Kitching \& Taylor, 2010) 
that marginalises over all functional perturbations about a
fixed function. This then provides goodness of fit that is the 
evidence for the fixed function over all other
functions. To achieve this we assume that the statistical properties of the data can be modelled by a multivariate 
Gaussian distribution. 
Using this `path-integral evidence' we also define a utility function
that includes a path-integral over the set of perturbed functions and
future data,
this can be used to calculate the expected value of evidence for a
fixed function over perturbations about that function for a future
experiment. 

We apply this methodology to the case of determining if 
cosmological experiments can determine the true dark energy equation of
state, as a function of redshift, $w(z)$, from perturbations about this
function. We use as a simple example an experiment that aims to 
measure $w(z)$ from changes in the Hubble parameter. 
This is a particularly important question because dark energy
accounts for the majority of the mass-energy content of the Universe
and its nature is entirely unknown; explanations include proposals 
to modify general relativity on cosmological scales and 
the addition of new fundamental fields, amongst others. The strongest 
prediction for $w(z)$ is that it has a value of $w(z)=-1$ for all redshifts, 
which would indicate that dark energy is a vacuum energy, or additional 
gravitational constant known as the `cosmological constant'. 
Measuring any deviation from this functional form, at any redshift, would
necessitate the need for new physics beyond the standard models of
cosmology and/or particle physics. This article is arranged as follows in Section \ref{Method} we present
the general methodology, in Section \ref{Results} we present an
application to dark energy experiments, in Section \ref{Conclusion}
we present conclusions. 

\vspace{-0.7cm}
\section{Method}
\label{Method}
We begin with a function that depends on a variable $x$ 
and some parameters $\theta$, that has a fixed (fiducial) functional 
form $f_F(x|\theta_0)$, where the parameters take particular values $\theta_0$.  
We perturb this such that any function could be written as a perturbation 
away from the fiducial 
\be 
f(x)=f_F(x|\theta_0)+\delta_f(x).
\ee
These functions exist in a set $M$ that contains the fiducial function, and 
all perturbations about that function. We can write a log-likelihood for the parameters $\theta$ in the case
that the data can be modelled by a Gaussian distribution
\be 
-2{\mathcal L}=\Delta D^T \C_0^{-1}\Delta D + \ln |\C_0|
\ee
where $\Delta D=\mu(\theta)-D$ is the difference of the mean, 
that is a function of the parameters $\theta$, and the data $D$ and we 
define a covariance $\C_0$.
We denote the Fisher matrix for the free parameters $\theta$ as 
$F_{\theta\theta}$ calculated using the covariance $\C_0$.

From Kitching \& Taylor (2011) we can write a likelihood for the parameters $\theta$ 
accounting for a path-integral marginalisation over the functional 
behaviour of $\delta_f(x)$, by changing the covariance such that 
\be 
\label{C1}
\C_M=[\C_0^{-1}-\C_0^{-1} {\mathcal F} \C_0^{-1}]^{-1}
\ee
where 
\be
\label{P}
{\mathcal F}=\int {\rm d}x{\rm d}x'\frac{\delta\mu}{\delta\delta_f(x)}
F^{-1}_{\delta_f\delta_f}(x,x')\frac{\delta\mu}{\delta\delta_f(x')}
\ee
the mean is now a function of $\theta$ and $\delta_f(x)$, and 
$F_{\delta_f\delta_f}(x,x')$ are the elements of the functional 
Fisher matrix for the perturbations $\delta_f$ 
marginalised over the parameters $\theta$. This is the `self-calibration' case discussed in 
Kitching \& Taylor (2011) where the prior in function-space is a flat top hat;  
for a detailed discussion of flat functional priors see Bornkamp (2011). 

In the case that a functional prior is included over the perturbed 
function-space $\delta_f(x)$ the covariance reduces (via the Woodbury
matrix identity) to 
\be
\label{C}
\C_M=\C_0+{\mathcal G}
\ee
where 
\be
\label{G}
{\mathcal G}=\int {\rm d}x{\rm d}x'\frac{\delta\mu}{\delta\delta_f(x)}
C_{P,\delta_f\delta_f}(x,x')\frac{\delta\mu}{\delta\delta_f(x')}
\ee
where $C_{P}$ is the covariance of a Gaussian functional prior. Either the flat or
Gaussian-prior cases (equations \ref{C1} and \ref{C}) can be used to construct a 
new Fisher matrix for the parameters 
$\theta$, now taking into account the marginalisation over
perturbations $\delta_f$ about the fixed function. As discussed in Taylor \& Kitching 
(2010) the flat prior case is different to a Gaussian prior with large variance, they have  
different functional weighting that means the latter 
allows for the use of the Woodbury identity whereas the former does not.

We can now extend the Bayesian evidence formalism of Taylor \& Kitching
(2010), to include the 
marginalisation over the perturbed functional behaviour. 
Referring to Taylor \& Kitching (2010, equation 51)
we write the evidence, assuming that the data can be modelled using Gaussian distributions, including marginalisation over all perturbations 
about the fiducial function   
\ba 
\ln E^M&=&\Delta D^T \C_M^{-1}\Delta D + \ln |\C_M| \nn
&+& 2\ln (V^M\sqrt{|F^M|}) - N_M\ln 2\pi
\ea
where $\C_M$ is defined either in equation (\ref{C}) or (\ref{C1}) depending on the 
prior. $V^M$ is the volume over which the prior distribution is defined. $F^M$ is the Fisher matrix for the fixed parameters after marginalisation over the pertubed case, 
this is defined as the Schur complement of the full Fisher matrix $F$ i.e. 
$F^M_{ij}= F_{ij}-F_{i\alpha}F^{-1}_{\alpha\beta}F_{\beta j}$, where $\alpha$ and $\beta$ here represent function 
elements of the total Fisher matrix, and Roman letters represent parameters of the fixed function. We can also write the
evidence for the fixed function by referring to Taylor \& Kitching (2010, equation 54)
\ba
\ln E^{0}&=&\Delta D^T \C_0^{-1}\Delta D + \ln |\C_0| \nn
&+& 2\ln (V^0\sqrt{|F^0|})-N_0\ln 2\pi,
\ea
where $\C_0$ is the covariance not marginalising over the perturbations 
$\delta_f(x)$, and $F^0$ is the part of the Fisher matrix for the parameters of the fixed function only. $V^0$ is the volume over which 
the prior distribution is defined, in the fixed 
function case. $N_0=N_D-N_P$ is the number of degrees of freedom 
where $N_D$ is the number of data points and $N_P$ is 
the number of free parameters. In the case of the perturbed case $N_M\rightarrow -\infty$. 
We address the infinity below. 

By combining the expressions for the evidence 
in the fixed plus perturbed ($F+P$) case $f(x)=f_F(x)+\delta_f(x)$, and 
the fixed case $f(x)=f_F(x)$ only ($F$), we can write a Bayes factor 
that quantifies the ability of the data to distinguish these models
\ba 
\label{B1}
{\mathcal B}_{F/(F+P)}&=&\ln E^0-\ln E^M\nn
&=&\Delta D^T (\C^{-1}_0-\C_M^{-1})\Delta D\nn
&+&\ln|\C_0 \C_M^{-1}|.
\ea
This is the evidence for the fixed model over the fixed plus the perturbed case. 

The final terms in the
evidence, the Occam factor, do not appear in this evidence ratio. This is 
because, in the perturbed case the parameter space volume tends to 
infinity. There are two motivations that could be used to justify the removal of these 
terms. Firstly the removal of the Occam term can be considered as a 
practical measure. The Bayes factor would then not 
account for the Occam factor, and is therefore not a true Bayesian measure, 
but does still capture the change to the posterior 
caused by the increase in covariance as a result of marginalisation over the larger 
parameter (function) space.

Alternatively 
it can be justified by modifying the prior volume $V^M$ in the perturbed case 
such that in the two different models this term cancels. 
This is achieved by scaling the volume $V^M$ 
(a hypersphere with radii scaled using the ratio of the expected Fisher matrices), over which we
allow the prior in the perturbed case to be defined, by 
the prior volume of the unperturbed case 
\be
V^M=V^0 (2\pi)^{(N_M-N_0)/2}\left[\frac{F^0}{F^M}\right]^{1/2}.
\ee
The above expression is the volume of the hypersphere $V^M$ scaled from $V^0$. 
Note that the integral over prior (outside this volume) can tend to infinity but that in this case, because we always 
scale with respect to $V^0$ 
in taking the evidence ratio, the Occam factor terms cancel to zero. 
This is equivalent to setting a prior on
the parameter volume, as a result this metric will not be sensitive to penalties induced from an unexpected
increase in volume. Without some justification for the removal of the Occam factor the evidence ratio would
result in an infinite result (the marginalised case having an infinite
number of degrees of freedom), but we note that the scaling is
convergent, and so asymptotically should behave correctly. 

\subsection{Expectation Value}
We now explore the expectation value of the above statistic, and show how we can 
forecast this Bayes factor for experimental optimisation.

For forecasting purposes we need to integrate over the future data vector, as 
described in Amara \& Kitching (2010) and 
Trotta et al. (2011) where a an `expected utility' or
`figure of merit' is defined as an integral over parameters (or functional integrals 
in our case), and data vectors. 
The utility function here case is the log Bayes factor for the
fixed model
over the perturbed case ${\mathcal B}_{F/(F+P)}$, given in equation
(\ref{B1}), where we have already performed the integrals over
functions and parameters. The final integral over the data vector 
arrives at an expression for the expected  
evidence ratio 
\ba
\label{BB}
\langle {\mathcal B}_{F/(F+P)}\rangle &=& 1-\Tr(\C_0\C_M^{-1})+\ln|\C_0\C_M^{-1}|. 
\ea
We recognise the two terms above as i) the fractional change in the covariance, 
ii) the fractional change in
the volume of the covariance due to the functional marginalisation. 
This expression now quantifies the expected ability of an experiment to
distinguish a fixed function $f_F(x)$ from the  
fixed plus perturbed case $f_F(x)+\delta_f(x)$. 

Interestingly, this in fact has a similar form to the 
\emph{Itakura-Saito distance} that is a measure of the 
`perceptual difference between an original spectrum $P$ and an
approximation $\hat{P}$ of that spectrum', of use in signal
processing, but here we derive this from first principles.  
More generally equation (\ref{BB}) is similar to the 
the Kullback-Leibler divergence $D_{\rm KL}(p1,p2)  = \int {\rm d}x
p_1(x) \ln (p_2(x)/p_1(x))$,  which is a measure of the difference 
between two distributions, in our case the probability of the true
function and the probability marginalised over all functions. The 
Kullback-Leibler divergence was investigated as a utility function or `figure 
of merit' for cosmology in Amara \& Refregier (2014). However this only looked at simple functional 
changes, while we generalise to the continuous functional case and derive from first 
principles.

We can also calculate the evidence for the fixed model over the
perturbed case   
\be
\langle {\mathcal B}_{F/P}\rangle=-\ln [{\rm e}^{-\langle {\mathcal B}_{F/(F+P)}\rangle}-1].
\ee
Here we 
assume that the evidences for the fixed and perturbed models are additive, $p(D|F+P) = p(D|F) + p(D|P)$, 
where the probability $p(D|F+P)$ corresponds 
to the `probability of the fixed OR perturbed' case (i.e. any function). 
This is true because $F$ and $P$ are mutually exclusive; where here we 
define the perturbed model-space as all functions \emph{except} the fiducial case (another way 
of formulating this is that fiducial function $F$ is a special case within the space of all functions, in that 
its prior volume is zero). This is an expression for the expectation value of the evidence for the 
true function over perturbations away from that true function. 
To calculate this we make the approximation that 
$\langle{\mathcal B}_{F/(F+P)}\rangle=\ln [\langle E^0/(E^0+E^M)\rangle]$. 

This statistic can be used to optimise experiments so that they will
have the ability to distinguish a fixed function from peturbations
about that function. In 
practice one would fit a free functional form to the data (see
e.g. Crittenden et al., 2011).
For calculation purposes one must assume a `fiducial' model for the
fixed function $f_F(x)$ around
which to calculate the Fisher matrices and covariances. However we
note that under the assumption that the data can be modelled by a multivariate 
Gaussian the expressions above are \emph{independent} of the choice of
fiducial function (see Taylor \& Kitching, 2010).

\vspace{-0.7cm}
\section{Example Application}
\label{Results}
We now apply this methodology to the dark energy equation of state $w(z)$,
where we have a model that includes a fixed function of $w_F(z)$
and we include perturbations about this model 
\be 
w(z)= w_F(z) + \delta_w(z),
\ee
where the function is over redshift $z$. 
\begin{figure*}
  \includegraphics[angle=0,clip=,width=\columnwidth]{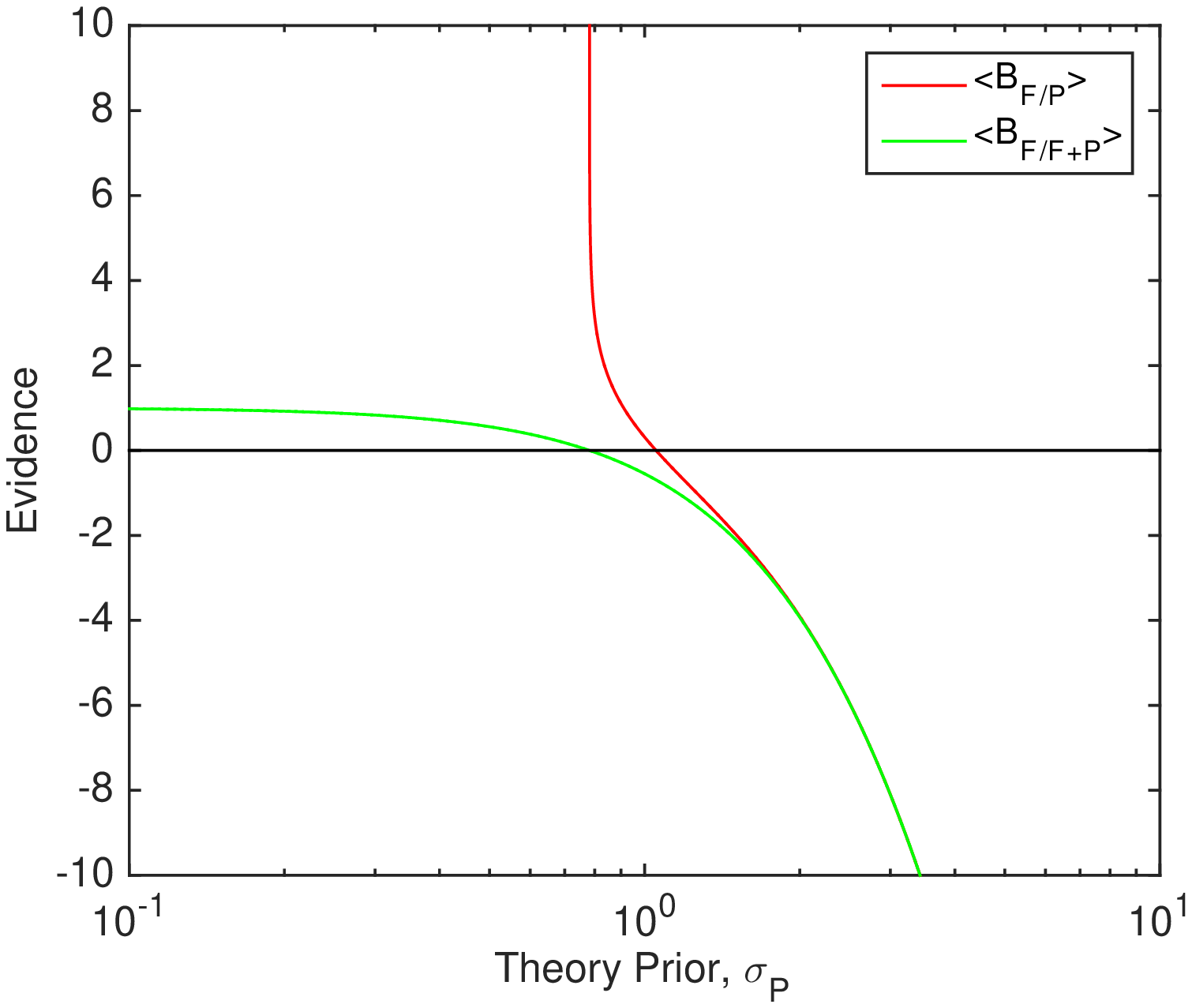}
  \includegraphics[angle=0,clip=,width=\columnwidth]{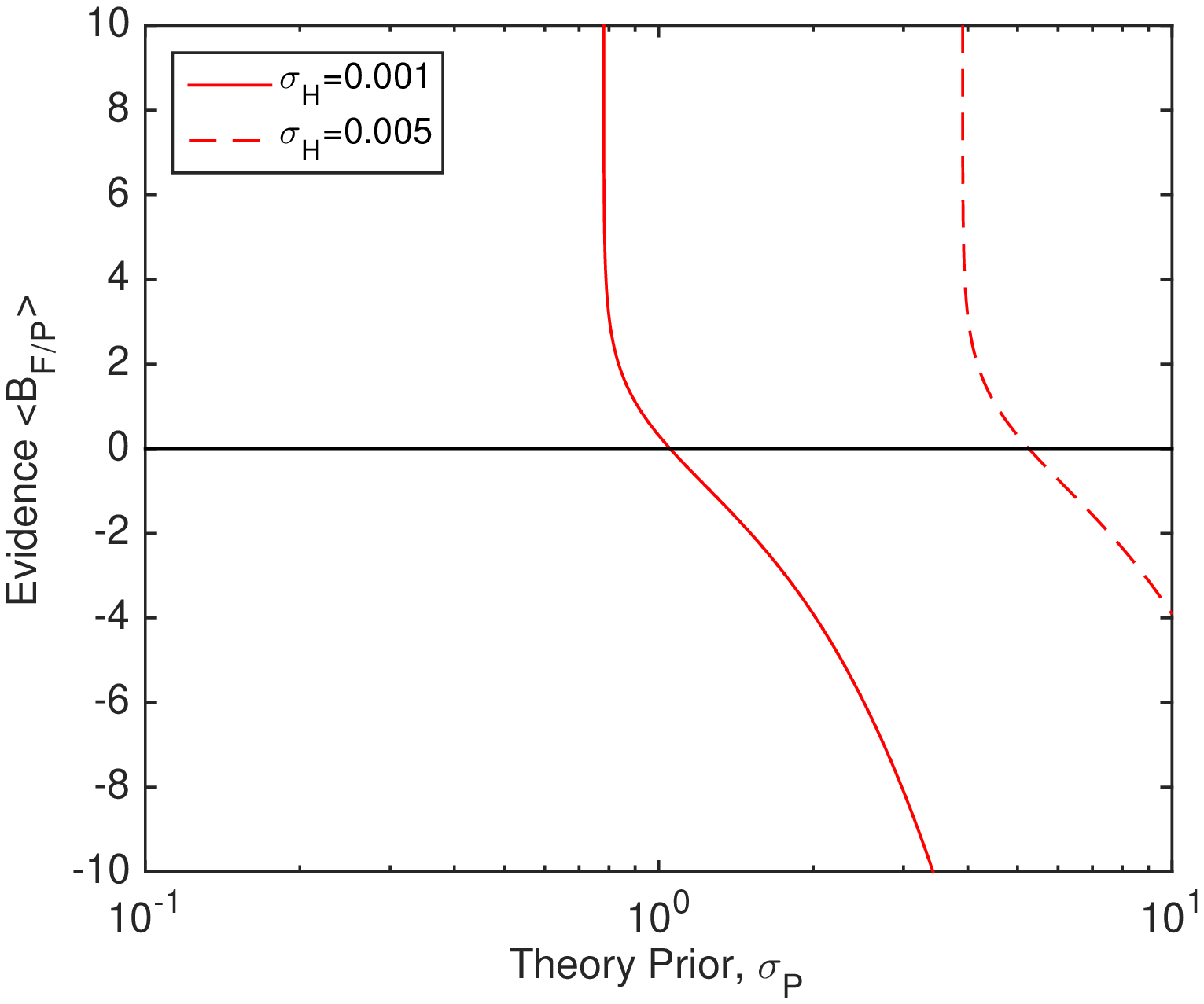}
 \caption{Left: The expected evidence $\langle{\mathcal B}_{F/P}\rangle$ (fixed vs. perturbed) and 
$\langle{\mathcal B}_{F/(F+P)}\rangle$ (fixed vs. all)
as a function of the pre-experimental theoretical prior. This is for the illustrative example of measuring a perturbation about a fixed value of the dark
energy equation of state $w(z)=-1$ using measurements of the Hubble parameter. Shown are expected evidence for an expected error on the measurements of
$\sigma_H=10^{-3}$. Right: The expected evidence $\langle{\mathcal B}_{F/P}\rangle$ (fixed vs. perturbed) for two different values of the expected error on $H$.}
 \label{w}
\end{figure*}
We look at an illustrative example in this paper as a proof of concept, where we consider 
a measure of the Hubble parameter $H(z)$ which is related to $w(z)$ via the follow equation 
\ba
H(z)&=&H_0[\Omega_M(1+z)^3\nn
&+&\Omega_X(1+z)^{3\int_0^z {\rm d}z' [1+w(z')]/(1+z')}]^{1/2}. 
\ea
where $H_0$ is the current rate of expansion, $\Omega_M$ and $\Omega_X$ are the dimensionless matter and 
dark energy densities respectively and $w(z)$ is the dark energy equation of state. We consider a hypothetical 
experiment that measures $H(z)$ in ten redshift bins between redshifts $0\leq z \leq 2$ with an accuracy of 
$\sigma_H$ in each redshift bin. We use a fiducial cosmology that assumes 
$H_0=70$kms$^{-1}$Mpc$^{-1}$, $\Omega_M=0.3$, 
$\Omega_X=0.7$ and $w_F(z)=-1$ $\forall$ $z$. 
We assume a functional prior on $w(z)$ that is diagonal in redshift 
$C_{P\delta_{w(z)}\delta_{w(z')}}=\sigma_P^2\delta^K_{zz'}$. In Figure \ref{w} we show an example where the 
measurement on $H(z)$ has an error bar of $\sigma_H=10^{-3}$, all errors have the same units as $H$. 

It is worth spending some time examining the generic behaviour seen in Figure \ref{w}. The theory prior is 
a functional prior around the fixed function. For the Bayes factor 
$\langle{\mathcal B}_{F/(F+P)}\rangle$ what we find is that as the theory prior approaches infinity the Bayes factor tends to minus 
infinity, this means that data would favour a perturbed model. As the theory prior tends to zero the 
Bayes factor $\langle{\mathcal B}_{F/(F+P)}\rangle\rightarrow 1$, as expected from equation (\ref{BB}), 
this means that if one performed this experiment with a small theoretical prior there would be no change in evidence 
$\langle{\mathcal B}_{F/(F+P)}\rangle$ i.e. 
with this prior \emph{it would not be worth doing this experiment} 
because the space of models tested is already well constrained by the prior, and the experiment is expected to favour the fixed function. 

The Bayes factor $\langle{\mathcal B}_{F/P}\rangle$ has an interesting and important behaviour. When 
$\langle{\mathcal B}_{F/(F+P)}\rangle=0$ there is a singularity where the expected evidence for 
the fixed model tends to infinity. This boundary is of particular importance. If one performs an experiment with a pre-experimental theoretical prior smaller 
than a particular value (leftward of this singularity in Figure \ref{w}) then the experiment is likely to return no change in evidence over that prior, and 
is likely to favour the fixed function. However, if one performs an experiment 
with a pre-experimental prior that is larger then the experiment is likely to return a change in evidence (either a weaker preference for the fixed function, 
or for the perturbed case). 
In colloquial terms this singularity demarks, for a given expected experimental error, 
and a given pre-experimental theoretical prior, whether the experiment is `worth doing'.

To explore this concept further we show in Figure \ref{w} the case of two different expected error bars on $H(z)$. Consider in this 
Figure the case that before doing the experiment the theoretical prior 
about $w(z)$ was $\sigma_P=1$, in this case it would be worth doing an experiment with an 
expected error of $\sigma_H=10^{-3}$ but it would not be worth doing an experiment with $\sigma_H=5\times 10^{-3}$. 

Furthermore as the error bar decreases the expected evidence for of finding a perturbation increases ($\langle {\mathcal B}_{F/P}\rangle$ becomes more 
negative). If one is rightward of the singularity in Figure \ref{w} then an experiment 
may be worth doing, but it can be highly likely that it will find no evidence for a perturbation. As the experimental error 
bar decreases one becomes more likely to find evidence for a perturbation over the fixed case (negative values of $\langle {\mathcal B}_{F/P}\rangle$).  
There is a point where the evidence of the fixed function becomes zero $\langle {\mathcal B}_{F/P}\rangle=0$, this occurs when $\langle {\mathcal B}_{F/(F+P)}\rangle=-\ln(2)$ i.e. that 
there is equal evidence for the fixed and perturbed case ($\langle {\mathcal B}_{F/(F+P)}\rangle=\ln[E_F/(E_F+E_P)]=\ln(1/2)$). 

\begin{figure}
  \includegraphics[angle=0,clip=,width=\columnwidth]{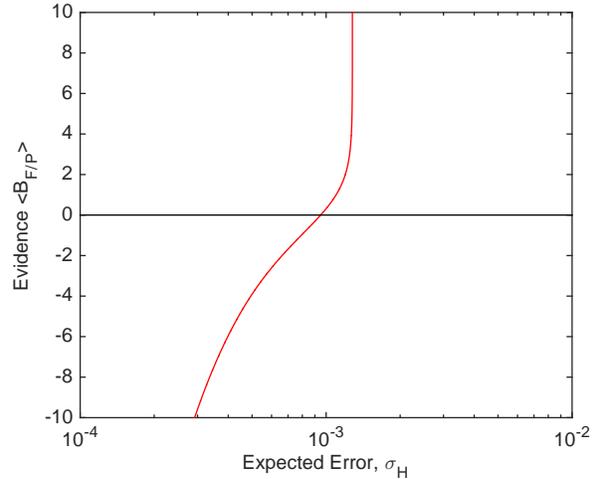}
 \caption{The expected evidence $\langle {\mathcal B}_{F/P}\rangle$ (fixed vs. perturbed) as a function of the expected error bar on the Hubble parameter $\sigma_H$; 
here we fix the pre-experimental theory prior to $\sigma_P=1$.}
 \label{h}
\end{figure}

We explore this in Figure \ref{h} where the expected evidence $\langle {\mathcal B}_{F/P}\rangle$ is shown as a function of the expected experimental error $\sigma_H$. When the 
expected error is larger there is no expected change in evidence due to the experiment. As the expected error decreases there is a regime where the experiment becomes 
worth doing - but the result is expected to confirm the non-perturbed model (i.e. given the theoretical prior the data is unlikely to provide evidence against the 
non-perturbed model). However when the expected error is small enough, in this case $\sigma_H\ls 10^{-3}$, the experiment is worth doing \emph{and} is 
expected to return evidence for a perturbed model over the non-perturbed case - if such a perturbation exists. 

\vspace{-0.7cm}
\section{Conclusion}
\label{Conclusion}
Here we present, under the assumption that the data can be modelled by a multivariate Gaussian distribution,  
a Bayesian goodness-of-fit which compares the evidence for a fixed functional form to the 
total evidence for all functions that are not the fixed functional
form. We then define a utility function using this metric that
marginalises over all functional perturbations about a fixed model. 

We apply this using an illustrative example of determining the dark energy equation state 
through measurements of the Hubble parameter. We find that the expected evidence provides a
clear metric, given a pre-experimental theoretical prior and expected error bar, 
that can quantify if an experiment is likely to produce a change in the evidence; answering the question 
is an experiment `worth doing'? We find that as the expected error bar for a planned experiment 
decreases the expected evidence for finding a perturbation can become significant, and we 
suggest that this metric may be useful in experimental design. For our illustrative example 
we show that in order for an experiment to expect to find strong evidence for a perturbation 
about $w(z)=-1$ a measurement of the Hubble parameter only would need to have errors of $\sigma_H\ls 10^{-3}$ over ten 
redshift bins between $0\leq z \leq 2$. 

This formalism should be applicable to any case of experimental design that seeks to maximise the 
expectation of finding perturbations about a functions, and as such can be used as a model-independent 
tool for experimental optimisation. 
\\
%__________________________________________________________________
\newpage
\noindent{\em Acknowledgements:} TDK is supported by a Royal Society Univeristy
Research Fellowship. We thank an anonymous referee for comments that improved an earlier version of this paper.

%__________________________________________________________________

%_________________________________________________________________

\onecolumn

\end{document}